\documentclass[runningheads]{llncs}

\usepackage{eccv}

\usepackage{eccvabbrv}
\usepackage{graphicx}
\usepackage{booktabs}
\usepackage{caption}
\usepackage[table]{xcolor}
\usepackage{wrapfig}
\definecolor{fullrowcolor}{RGB}{200,200,200}
\definecolor{oursrowcolor}{RGB}{230,230,230}
\usepackage{algorithm}
\usepackage{algorithmic}

\usepackage[accsupp]{axessibility}  

\usepackage{hyperref}

\usepackage{orcidlink}

\begin{document}

\title{DASH: Dynamic Audio-Driven Semantic Chunking for Efficient Omnimodal Token Compression}

\titlerunning{DASH}

\author{Bingzhou Li\inst{1,2} \and Tao Huang\inst{1}\thanks{Correspondence to: Tao Huang (\url{t.huang@sjtu.edu.cn}).}}

\authorrunning{B.~Li and T.~Huang}

\institute{Shanghai Jiao Tong University, Shanghai, China \and
Tongji University, Shanghai, China}

\maketitle

\begin{abstract}
Omnimodal large language models (OmniLLMs) jointly process audio and visual streams, but the resulting long multimodal token sequences make inference prohibitively expensive. Existing compression methods typically rely on fixed window partitioning and attention-based pruning, which overlook the piecewise semantic structure of audio-visual signals and become fragile under aggressive token reduction. We propose Dynamic Audio-driven Semantic cHunking (DASH), a training-free framework that aligns token compression with semantic structure. DASH treats audio embeddings as a semantic anchor and detects boundary candidates via cosine-similarity discontinuities, inducing dynamic, variable-length segments that approximate the underlying piecewise-coherent organization of the sequence. These boundaries are projected onto video tokens as a soft temporally co-registered segmentation prior. Within each segment, token retention is determined by a tri-signal importance estimator that fuses structural boundary cues, representational distinctiveness, and attention-based salience, mitigating the sparsity bias of attention-only selection. This structure-aware allocation preserves transition-critical tokens while reducing redundant regions. Extensive experiments on AVUT, VideoMME, and WorldSense demonstrate that DASH maintains competitive or superior accuracy while achieving higher compression ratios compared to prior methods. Code is available at: \url{https://github.com/laychou666/DASH}.
\end{abstract}

\section{Introduction}
\label{sec:intro}

Recent multimodal and omnimodal large language models~\cite{xu2025qwen25omni, team2023gemini, achiam2023gpt4, fu2024vita, shu2025avllm, tang2025videosalmonn2} extend video-language systems~\cite{zhang2023videollama, li2023videochat, zhang2024llavavideo, chen2024internvl, liu2023llava, lin2023videollava} toward unified processing of visual, audio, and textual signals. By integrating speech, scene dynamics, and textual reasoning, these models enable richer multimodal understanding across tasks such as video question answering and real-world reasoning. However, this capability comes at a significant computational cost: a single video clip can produce tens of thousands of multimodal tokens, and the quadratic complexity of self-attention quickly makes inference memory- and latency-bound.

Token compression~\cite{chen2024fastv, shang2025llavaprumerge, bolya2022tome, yang2025visionzip, xing2025pyramiddrop, ye2025fitprune} has become a practical solution to mitigate this bottleneck. By pruning or merging tokens before they enter the language model, prior work reduces sequence length without retraining large models. Yet despite steady progress, existing approaches share a critical assumption: multimodal tokens are treated as flat, uniformly structured sequences. Compression decisions are typically made within fixed windows~\cite{tao2025omnizip,bolya2022tome,chen2025streamingtom} and guided primarily by attention scores. This positional partitioning neglects the piecewise-coherent structure of audio-visual sequences, where semantic transitions correspond to distributional shifts in embedding space.

The core challenge lies in the intrinsic structure of audio-visual signals. Unlike static images, audio and video streams evolve over time with highly \textbf{non-uniform} semantic density. Speech contains natural boundaries (\emph{e.g.}, pauses, topic transitions, and speaker shifts) that often provide useful cues for meaningful visual changes. These transitions delineate coherent semantic units that should ideally be preserved as atomic structures during compression. When tokens are grouped uniformly and selected via sparse attention alone, these structural transitions are fragmented or discarded, leading to disproportionate information loss under aggressive pruning. In other words, current compression strategies optimize local redundancy while overlooking global semantic organization. From this perspective, token compression can be viewed as allocating limited representational capacity across a temporally structured and semantically segmented signal.

\begin{figure}[t]
\centering
\includegraphics[width=\linewidth]{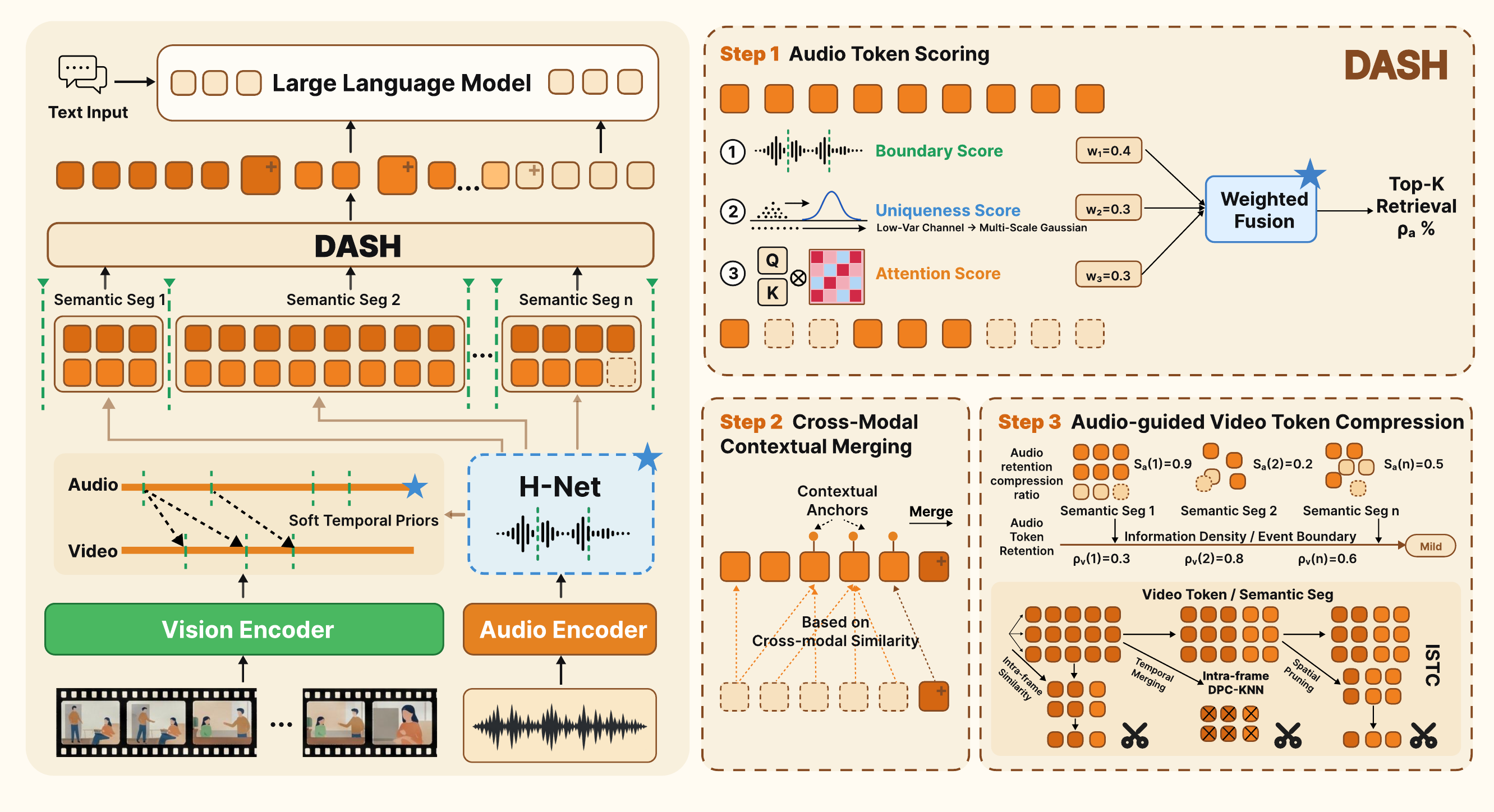}
\caption{\textbf{Overview of DASH.} \textbf{Left}: Overall DASH pipeline, where audio/video inputs are encoded, audio semantic boundaries are detected, and projected as soft temporal priors to form dynamic audio-guided segments before compression. \textbf{Right}: DASH module details, including tri-signal audio token scoring, contextual merging, and audio-guided video token compression with mild boundary-neighborhood retention.}
\label{fig:framework}
\end{figure}

In this work, we argue that effective omnimodal compression should be guided by semantic structure rather than positional regularity. In particular, the audio stream provides a natural structural prior. Compared to visual tokens, audio embeddings exhibit sharper distributional discontinuities at semantic transitions (e.g., pauses and topic shifts), making them a more reliable boundary signal. Because audio and video inputs are temporally co-registered, these audio transitions provide useful cues for organizing video tokens without requiring exact audio-visual boundary equality. Audio therefore acts as a semantic anchor that can guide both segmentation and importance allocation across modalities.

Building on this perspective, we propose \textbf{DASH (Dynamic Audio-driven Semantic cHunking)}, a training-free framework for structure-aware omnimodal token compression (\cref{fig:framework}). DASH models the multimodal sequence as a set of dynamically inferred semantic segments instead of imposing fixed windows. Specifically, it detects boundary candidates from cosine-similarity drops between adjacent audio embeddings, treating sharp representational changes as proxies for semantic discontinuities. These boundaries induce variable-length chunks that respect content rhythm and approximate the underlying piecewise structure of the sequence.

To introduce a cross-modal structural prior, detected audio boundaries are projected onto video token indices via temporal ratio mapping, yielding audio-guided video segments without additional learning. Within each segment, token selection becomes a \emph{multi-signal importance estimation} problem. We combine (i) boundary probability as a structural prior, (ii) multi-scale Gaussian uniqueness capturing representational distinctiveness, and (iii) normalized attention reflecting model-perceived salience. Their fusion yields a smoother importance landscape (\cref{fig:token_heatmap}) and alleviates the heavy-tailed sparsity of attention-only selection.

DASH introduces no learnable parameters and operates as a lightweight plug-in between modality encoders and the LLM. By reallocating compression capacity according to semantic density and preserving transition-critical tokens, DASH maintains narrative continuity and cross-modal coherence under aggressive pruning. Experiments on AVUT~\cite{yang2025avut}, VideoMME~\cite{fu2025videomme}, and WorldSense~\cite{hong2025worldsense} with Qwen2.5-Omni (7B and 3B) show that DASH maintains competitive accuracy at 25\% token retention, matching or exceeding prior methods operating at 35\%, while achieving up to 3.8$\times$ prefill speedup and 1.7$\times$ end-to-end latency reduction.

Our contributions are threefold.
\begin{itemize}
    \item We formulate omnimodal token compression as a structure-aware segmentation problem, showing that compression should follow semantic boundaries rather than fixed positional windows.
    \item We propose DASH, a training-free, audio-anchored framework that detects semantic discontinuities in embedding space and uses them as soft temporal priors for video segmentation and density-aware compression.
    \item We demonstrate on three audio-video benchmarks that structure-aware compression enables stable performance at substantially lower token retention (25\%) than prior methods (35\%), achieving significant speedups with negligible overhead.
\end{itemize}

\section{Related Work}

\subsection{Token Compression for Multimodal LLMs}

Token compression reduces input tokens to alleviate the computational bottleneck of LLM inference, exploiting redundancy in multimodal inputs.
For \emph{image} inputs, FastV~\cite{chen2024fastv} performs training-free inference acceleration by monitoring attention patterns at a designated LLM layer and pruning uninformative visual tokens during prefill.
LLaVA-PruMerge~\cite{shang2025llavaprumerge} selects representative tokens via adaptive clustering and absorbs the rest through weighted averaging.
ToMe~\cite{bolya2022tome} introduces a training-free bipartite matching algorithm that progressively merges similar token pairs across ViT layers.
Other methods use attention-based selection~\cite{yang2025visionzip}, progressive layer-wise reduction~\cite{xing2025pyramiddrop,ye2025fitprune}, or learned projectors~\cite{li2025tokenpacker}.
For \emph{video} inputs, DyCoke~\cite{tao2025dycoke} proposes a two-stage pipeline separating temporal and spatial redundancy, while others apply spatiotemporal merging~\cite{shao2025holitom,shen2025fastvid} or attention-based pruning~\cite{huang2025prunevid,shen2025longvu} (see~\cite{shao2025survey} for a survey).
However, these methods operate within single modalities and ignore cross-modal correlations.
OmniZip~\cite{tao2025omnizip} pioneers \emph{omnimodal} compression using audio attention to guide video pruning with interleaved spatio-temporal compression.
We improve upon OmniZip with dynamic semantic chunking, audio-guided temporal priors for video segmentation, and tri-signal fusion scoring.

\subsection{Dynamic Chunking and Boundary Detection}

Traditional sequence processing relies on fixed-size segmentation: subword tokenizers (\eg, BPE, WordPiece) split text at predetermined granularity, and token compression methods uniformly group frames into equal-sized windows.
This uniform strategy ignores semantic structure---fragmenting coherent units across arbitrary boundaries and failing to adapt to information density.
Dynamic chunking addresses this by adapting segment lengths to semantic structure, yielding variable-length chunks that respect natural boundaries.
MEGABYTE~\cite{yu2023megabyte} models byte sequences with fixed-size patches in a multiscale architecture, while Late Chunking~\cite{gunther2024late} segments text for contextual retrieval.
H-Net~\cite{hnet2025} learns dynamic chunking boundaries end-to-end using cosine-similarity-based routing between adjacent representations.
MMHNet~\cite{simon2026echoes} further applies hierarchical routing and dynamic chunking to long-form video-to-audio generation.
In video and audio, boundary detection has been studied for shot/scene segmentation~\cite{shotdetect,sceneseg}, action localization~\cite{actionloc}, and token pruning~\cite{lin2025speechprune,li2023adjacent}, but these typically require task-specific supervision or operate on raw signals.
Inspired by H-Net's cosine-similarity routing for dynamic chunking, we use adjacent-representation similarity as a training-free boundary cue for audio tokens in OmniLLMs, and extend it to audio-guided video segmentation by projecting audio boundaries onto video indices as soft temporal priors.

\section{Method}
\label{sec:method}

  We present DASH, a training-free framework that operationalizes the principle of structure-aware compression through three key innovations. As illustrated in \cref{fig:framework}, given an input video with audio, DASH first detects semantic boundaries in the audio stream via cosine-similarity-based boundary detection (\cref{sec:boundary}). It then projects these boundaries onto video token indices through temporal ratio mapping to form dynamic audio-guided segments (\cref{sec:audiodriven}), providing a soft cross-modal structural prior without learned parameters. Finally, it scores tokens within each segment via tri-signal fusion for selective retention (\cref{sec:trisignal}). The retained tokens are passed to the LLM for inference.

\subsection{Problem Formulation}
\label{sec:problem}

Given an OmniLLM with audio encoder $\mathcal{E}_a$ and video encoder $\mathcal{E}_v$, we obtain audio token sequence $\mathbf{A} = \{a_t\}_{t=1}^{N_a} \in \mathbb{R}^{N_a \times D}$ and video token sequence $\mathbf{V} = \{v_t\}_{t=1}^{N_v} \in \mathbb{R}^{N_v \times D}$, where $D$ is the embedding dimension. For video, we have $N_v = F \times K$ tokens, where $F$ is the number of frames and $K$ is the number of tokens per frame.

\textbf{Static grouping in prior work.} Existing methods (\eg, OmniZip~\cite{tao2025omnizip}) divide video into fixed groups of 4 frames ($G_v = 4K$ tokens) and audio into fixed groups of 50 tokens ($G_a = 50$), regardless of semantic content. This rigid partitioning treats information-dense segments (rapid dialogue, complex actions) and information-sparse ones (silence, static scenes) identically.

\textbf{Our goal.} We aim to replace static grouping with dynamic semantic chunking that adapts to content structure, and replace single-signal token selection with multi-signal fusion scoring that captures structural, content, and model-perceived importance jointly.

\subsection{Dynamic Semantic Chunking (Segment-Level)}
\label{sec:boundary}

Rather than imposing fixed boundaries every $G_a$ tokens, we let the audio content itself determine where segmentation should occur. This design choice directly addresses the core limitation of prior work: \emph{positional partitioning treats information-dense and information-sparse segments identically}. Our approach is motivated by a key property of audio encoder embeddings: within a semantically coherent segment---such as a continuous sentence or a single topic---adjacent tokens encode similar contextual information, producing high cosine similarity. Conversely, at semantic transitions---sentence pauses, topic shifts, or speaker changes---the embedding space shifts abruptly, causing a sharp similarity drop. This property provides a training-free signal for boundary detection that is both reliable and computationally efficient. Inspired by H-Net's~\cite{hnet2025} cosine-similarity routing for dynamic chunking, we adapt adjacent-representation similarity to audio tokens as a training-free boundary cue.

\noindent\textbf{Boundary probability computation.}
For the audio token sequence $\mathbf{A} = \{a_t\}_{t=1}^{N_a}$, we compute the boundary probability at position $t$ as:
\begin{align}
  \text{sim}_t &= \frac{\langle a_{t-1}, a_t \rangle}{\|a_{t-1}\| \cdot \|a_t\|}, \label{eq:cossim} \\
  p_t^{\text{boundary}} &= \text{clip}\left(\frac{1 - \text{sim}_t}{2}, 0, 1\right), \label{eq:boundaryprob}
\end{align}
where $p_t^{\text{boundary}} \in [0, 1]$ is high when adjacent tokens are dissimilar (semantic discontinuity) and low when they are similar (semantic continuity). The first token is always treated as a boundary ($p_1^{\text{boundary}} = 1$).

\noindent\textbf{Boundary detection with minimum chunk constraint.}
A boundary is detected at position $t$ when the cosine similarity drops below threshold $\tau_a$ (default 0.4) \emph{and} the distance from the last boundary exceeds a minimum chunk size $C_{\min}$ (default 30 tokens, $\approx$1 second of audio):
\begin{equation}
  m_t^{\text{boundary}} = \mathbb{I}(\text{sim}_t < \tau_a) \cdot \mathbb{I}(t - t_{\text{last}} \geq C_{\min}),
  \label{eq:boundarymask}
\end{equation}
where $t_{\text{last}}$ is the position of the most recent boundary. The minimum chunk constraint is essential: without it, noise-induced similarity fluctuations would produce excessively short chunks (2--3 tokens) that are too small for meaningful importance comparison. The default $C_{\min} = 30$ corresponds to approximately 1 second of audio, ensuring each chunk spans at least one prosodic unit while still capturing sentence-level pauses and topic shifts. This produces a set of audio boundary positions $\mathcal{B}_a = \{b_0^a, b_1^a, \ldots, b_S^a\}$.

The resulting variable-length chunks ensure that tokens within each chunk share coherent semantics for fairer importance comparison, and naturally adapt to information density.

\subsection{Audio-Guided Video Segmentation (Cross-Modal Level)}
\label{sec:audiodriven}

A key observation in audio-guided video understanding is that \emph{speech is a strong semantic carrier}: audio boundaries (sentence pauses, topic transitions) often provide useful cues for visual semantic transitions (scene changes, action shifts). We do not assume exact audio-visual boundary coincidence. Instead, since audio and video tokens are temporally co-registered in the OmniLLM's time-window structure, temporal position provides a practical proxy for transferring audio-derived structure to video tokens. We therefore use audio boundaries as a soft temporal prior for video segmentation. This replaces rigid fixed-step grouping with content-aware segments without adding learned parameters.

\noindent\textbf{Temporal ratio mapping.}
Given audio boundaries $\mathcal{B}_a$ detected on $N_a$ audio tokens, we project them onto video token indices via linear scaling:
\begin{equation}
  b_i^v = \left\lfloor b_i^a \cdot \frac{N_v}{N_a} \right\rfloor, \quad i = 0, 1, \ldots, S,
  \label{eq:mapping}
\end{equation}
where $N_v$ is the total number of video tokens. The projected boundaries $\mathcal{B}_v = \{b_0^v, b_1^v, \ldots, b_S^v\}$ are deduplicated, sorted, and clamped to $[0, N_v]$, with $b_0^v = 0$ and $b_S^v = N_v$ enforced.
The resulting segments are audio-guided rather than hard visual cuts: they encourage video compression to follow audio-derived temporal structure, while the final video token mask is still computed from visual embeddings within each segment.

\noindent\textbf{Strength-based boundary refinement.}
The projected boundaries $\mathcal{B}_v$ may produce video segments shorter than the minimum required by interleaved spatial-temporal compression ($2K$ tokens, i.e., two frames). To resolve this, we apply a greedy refinement: each inner boundary $b_i^v$ is assigned a strength $p_{b_i^a}^{\text{boundary}}$ from its corresponding audio position. We sort inner boundaries by strength in descending order and greedily insert them into the final set $\mathcal{B}_v^*$, initialized as $\{0, N_v\}$. A boundary is accepted only if both adjacent segments exceed $2K$ tokens:
\begin{equation}
  b_i^v \in \mathcal{B}_v^* \iff (b_i^v - b_{\text{left}}) \geq 2K \;\wedge\; (b_{\text{right}} - b_i^v) \geq 2K,
  \label{eq:refinement}
\end{equation}
where $b_{\text{left}}$ and $b_{\text{right}}$ are the nearest existing boundaries in $\mathcal{B}_v^*$. By prioritizing the strongest audio-indicated boundaries (highest $p^{\text{boundary}}$) rather than arbitrarily dropping boundaries, this greedy strategy preserves the most reliable temporal priors while ensuring all segments are large enough for the interleaved spatial-temporal compression that follows. DASH compresses each segment independently from the original encoder embeddings; compressed outputs from earlier segments are not used to determine later boundaries or later token features.

\subsection{Tri-Signal Fusion Token Scoring (Token-Level)}
\label{sec:trisignal}

\begin{figure}[!t]
\centering
\includegraphics[width=\linewidth]{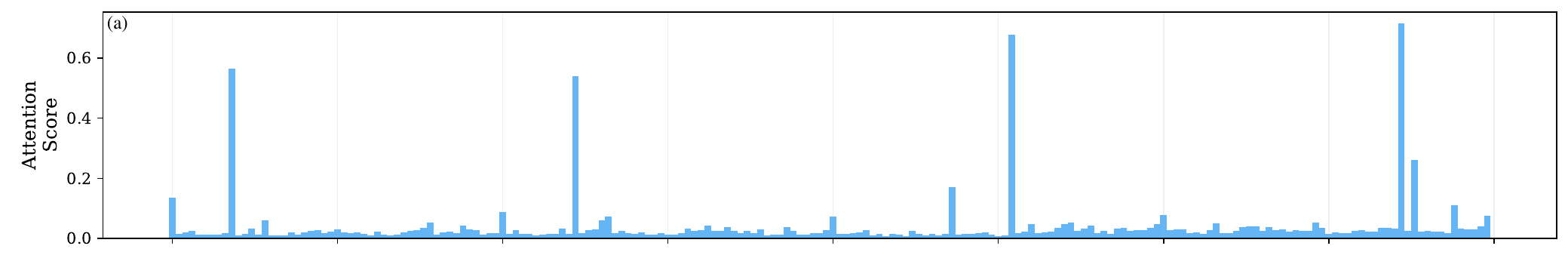}\\[1pt]
\includegraphics[width=\linewidth]{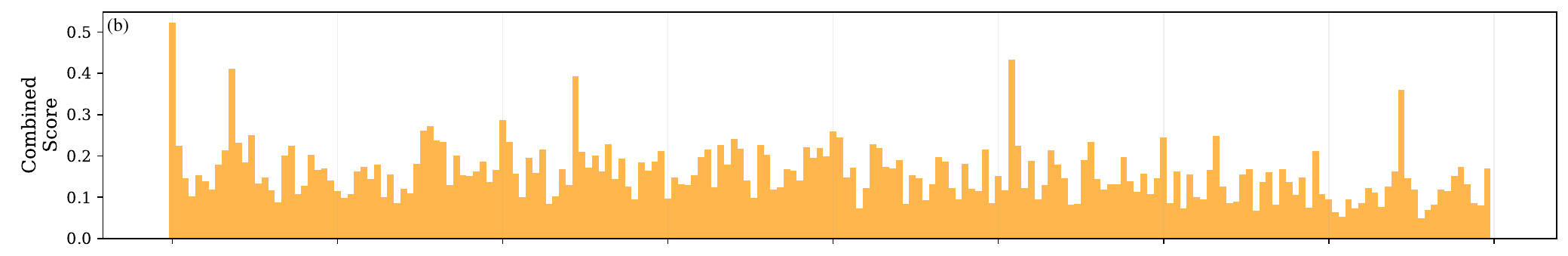}\\[1pt]
\includegraphics[width=\linewidth]{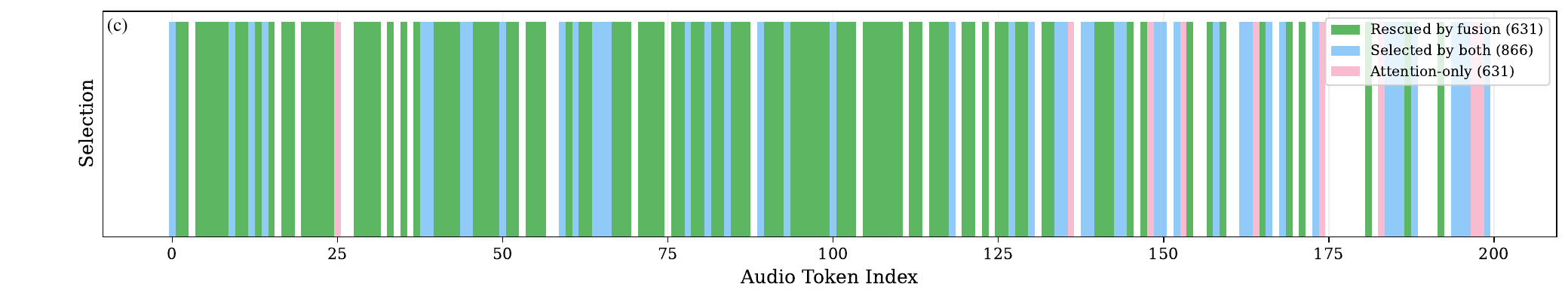}
\caption{\textbf{Token scoring and selection comparison.} (a)~Attention score alone is extremely sparse, concentrating on a few tokens. (b)~Tri-signal combined score produces a balanced importance landscape. (c)~Selection comparison: green = rescued by fusion (631), blue = selected by both (866), pink = attention-only tokens replaced (631). Fusion rescues $\sim$42\% of retained tokens that attention alone would discard.}
\label{fig:token_heatmap}
\end{figure}

Existing methods select audio tokens solely by attention scores from the audio encoder. However, as shown in \cref{fig:token_heatmap} (a), attention distributions are extremely sparse---importance concentrates on a handful of tokens while the vast majority receive near-zero scores. This sparsity is problematic for compression: a single signal cannot capture the full spectrum of token importance. Structurally critical tokens at semantic boundaries and content-distinctive tokens carrying unique information are discarded if they happen to fall outside the attention spotlight. The consequence is that aggressive compression (e.g., 25\% retention) based on attention alone destroys narrative continuity by removing transition anchors. To address this fundamental limitation, we propose a tri-signal fusion mechanism that combines three complementary importance signals, each capturing a different aspect of token importance: structural criticality, content distinctiveness, and model-perceived salience.

\noindent\textbf{Signal 1: Boundary probability ($w_b = 0.4$).}
Tokens at semantic boundaries are content transition points and should be prioritized. We normalize the boundary probability from \cref{eq:boundaryprob}
\begin{equation}
  s_t^{\text{bnd}} = \frac{p_t^{\text{boundary}}}{\max_j p_j^{\text{boundary}} + \epsilon}.
  \label{eq:signal_boundary}
\end{equation}

\noindent\textbf{Signal 2: Probabilistic density-based uniqueness ($w_u = 0.3$).}
Inspired by the density-peak clustering (DPC) principle used in VidCom~\cite{liu2025vidcom,dpcknn}, we measure the distinctiveness of each token. Unlike vanilla DPC which uses a hard distance cutoff for density estimation~\cite{dpcknn}, we propose a multi-scale Gaussian kernel to capture token importance across various feature granularities, providing a continuous and robust density approximation.

To enhance robustness, we first perform \textbf{low-variance channel selection}. Given token features $\mathbf{A} \in \mathbb{R}^{N_a \times D}$, we compute per-channel variance $\sigma_d^2 = \text{Var}([\mathbf{A}]_{:,d})$ and retain the bottom-$\lfloor D/2 \rfloor$ channels by variance, yielding $\tilde{\mathbf{A}} \in \mathbb{R}^{N_a \times D/2}$. This preprocessing step ensures that our uniqueness score is computed based on stable semantic dimensions rather than noisy, high-variance channels that may contain transient artifacts.

We then compute the $\ell_2$-normalized features $\hat{\mathbf{A}} = \text{normalize}(\tilde{\mathbf{A}})$ and the global center $\mathbf{c} = \frac{1}{N_a}\sum_t \hat{a}_t$. To approximate the local density $\rho$ defined in DPC-KNN~\cite{dpcknn} in a continuous space, we define the multi-scale Gaussian similarity:
\begin{equation}
  g_t = \sum_{\alpha \in \mathcal{A}} \exp\left(-\frac{\|\hat{a}_t - \mathbf{c}\|^2}{2\alpha}\right),
  \label{eq:gaussian}
\end{equation}
where $\mathcal{A} = \{0.125, 0.25, 0.5, 1.0, 2.0\}$ are multi-scale bandwidth parameters that enable the kernel to capture both fine-grained and coarse-grained feature variations. The uniqueness score is:
\begin{equation}
  s_t^{\text{uniq}} = 1 - \frac{g_t}{\max_j g_j + \epsilon}.
  \label{eq:uniqueness}
\end{equation}

\noindent\textbf{Signal 3: Attention score ($w_a = 0.3$).}
We use the attention scores from the audio encoder, normalized to $[0, 1]$:
\begin{equation}
  s_t^{\text{attn}} = \frac{\text{attn}_t}{\max_j \text{attn}_j + \epsilon}.
  \label{eq:signal_attn}
\end{equation}

\noindent\textbf{Fusion and selection.}
The final importance score is the weighted sum:
\begin{equation}
  s_t = w_b \cdot s_t^{\text{bnd}} + w_u \cdot s_t^{\text{uniq}} + w_a \cdot s_t^{\text{attn}},
  \label{eq:fusion}
\end{equation}
where $w_b = 0.4$, $w_u = 0.3$, $w_a = 0.3$. We select the top-$N_{\text{keep}}$ tokens by $s_t$ as the retained audio tokens, where $N_{\text{keep}} = \lfloor (1 - \rho_a) \cdot N_a \rfloor$.

The three signals capture orthogonal importance dimensions: boundary probability measures \emph{structural} importance at transition points, density-based uniqueness measures \emph{content} distinctiveness, and attention reflects \emph{model-perceived} importance. A boundary token may receive low attention but high boundary probability; fusion ensures such tokens are not overlooked. We set $w_b = 0.4$ slightly higher because structural boundaries are critical under aggressive compression. When boundary information is unavailable (\eg, very short sequences), the method falls back to attention-only selection.

\subsection{Boundary-Aware Video Compression}
\label{sec:compression}

Given the audio-guided video segments, we perform adaptive compression within each segment. The key idea is that not all segments deserve equal compression: segments associated with information-dense audio (rapid speech, complex narration) should retain more video tokens to preserve the rich visual context, while segments associated with silence or ambient noise can be compressed more aggressively.

\noindent\textbf{Audio-guided adaptive compression ratio.}
We use the audio retention rate as a proxy for segment-level information density. For each video segment $s$ associated with audio interval $[b_{s-1}^a, b_s^a)$, we compute the audio retention rate $\bar{m}_a^{(s)}$ (fraction of audio tokens retained in that segment by tri-signal fusion). The video compression ratio is then adapted:
\begin{equation}
  \rho_v^{(s)} = \rho_v + \lambda_r (0.5 - \bar{m}_a^{(s)}), \quad \rho_v^{(s)} \in [0.1, 0.95],
  \label{eq:adaptiveratio}
\end{equation}
where $\rho_v$ is the base video compression ratio and $\lambda_r = 0.1$ controls adaptation strength. Segments with high audio retention (semantically important) receive lower compression; segments with low audio retention receive higher compression.

\noindent\textbf{Boundary-neighborhood retention.}
For frames near projected boundary positions, we mildly increase the retention ratio:
\begin{equation}
  r_f^{\text{boundary}} = r_s + (1 - r_s) \cdot 0.3 \cdot p_f^{\text{boundary}},
  \label{eq:boundaryprotect}
\end{equation}
where $r_s = 1 - \rho_v^{(s)}$ is the base retention ratio. This is a conservative safeguard, not a hard assumption that the projected audio boundary is an exact visual cut. It preserves transition-neighboring visual context while leaving the final video token choice to visual-feature-based pruning.

\noindent\textbf{Interleaved spatial-temporal pruning.}
Within each segment, we adopt the interleaved spatial-temporal compression (ISTC) strategy~\cite{tao2025omnizip} with the adaptive retention ratio: \textbf{Even frames}: Spatial pruning via DPC-KNN~\cite{dpcknn} that removes tokens with high local density (spatially redundant). \textbf{Odd frames}: Temporal pruning that removes tokens most similar to the previous frame (temporally redundant).

\section{Experiments}
\label{sec:experiments}

\subsection{Evaluation Setups and Implementation Details}
\label{sec:setup}

\noindent\textbf{Benchmarks.}
Following OmniZip~\cite{tao2025omnizip}, we evaluate on established audio-video understanding benchmarks: AVUT~\cite{yang2025avut}, VideoMME~\cite{fu2025videomme}, and WorldSense~\cite{hong2025worldsense}. AVUT is an audio-centric video understanding benchmark focusing on six tasks: event localization (EL), object matching (OM), OCR matching (OR), information extraction (IE), content counting (CC), and character matching (CM). VideoMME is widely used for video-understanding evaluations where including audio can improve accuracy. WorldSense assesses models' ability to understand audio and video jointly across eight domains.

\noindent\textbf{Comparison methods.}
Given the absence of token pruning methods specifically designed for the omnimodal setting, we follow OmniZip and select representative prior methods from single-modal domains for adaptation: FastV~\cite{chen2024fastv} performs training-free inference-time pruning by utilizing the attention score matrix of the $L$-th layer; DyCoke~\cite{tao2025dycoke} applies its TTM module to both video and audio tokens; Random pruning serves as a control group. We also compare directly with OmniZip~\cite{tao2025omnizip}. For fair comparison, we reproduce results for OmniZip and DASH, while results for Random, FastV, and DyCoke are taken from OmniZip~\cite{tao2025omnizip}.

\begin{table*}[t]
\centering
\caption{\textbf{Comparison of different methods on omnimodal (audio \& video) QA benchmarks.} Norm.\ Avg.\ is the mean of per-benchmark normalized scores, where the baseline accuracy is 100\%. ``-'' indicates OOM error. $^\dagger$FastV Norm.\ Avg.\ is computed from AVUT only. Best result among token pruning methods is in \textbf{bold}, second best is \underline{underlined}.}
\label{tab:main}
\resizebox{\textwidth}{!}{
\begin{tabular}{l|cc|ccccccc|c|c}
\toprule
Method & Retained & FLOPs & & & & AVUT & & & & VideoMME & Norm.  \\
 & Ratio & Ratio & EL & OR & OM & IE & CC & CM & Avg. & wo & Avg. \\
\midrule
\multicolumn{12}{l}{\textbf{Qwen2.5-Omni-7B}} \\
\midrule
\rowcolor{fullrowcolor} Full Tokens & 100\% & 100\% & 38.2 & 67.8 & 59.6 & 85.6 & 44.1 & 66.7 & 64.5 & 66.0 & 100\% \\
Random & 40\% & 34\% & 31.7 & 58.5 & 53.3 & 74.9 & \textbf{43.2} & 59.0 & 56.9 & 65.0 & 93.4\% \\
FastV & 35\% & 42\% & 24.1 & 60.7 & 54.3 & 81.6 & \underline{40.7} & 58.3 & 57.8 & - & 89.6\%$^\dagger$ \\
DyCoke (V\&A) & 35\% & 29\% & 32.9 & 62.1 & 54.9 & 74.5 & 39.0 & 58.3 & 57.4 & 65.2 & 93.9\% \\
OmniZip & 35\% & 29\% & \underline{33.3} & \textbf{67.2} & 54.6 & 84.7 & 38.4 & \underline{61.4} & 60.6 & 66.0 & 97.0\% \\
\rowcolor{oursrowcolor} DASH (ours) & 35\% & 29\% & 32.0 & 62.1 & \textbf{58.9} & \underline{85.5} & 40.4 & \textbf{64.2} & \textbf{61.5} & \textbf{66.7} & \textbf{98.2\%} \\
\rowcolor{oursrowcolor} DASH (ours) & 25\%  & 20\% & \textbf{36.0} & \underline{64.7} & \underline{56.0} & \textbf{86.3} & 36.5 & 60.8 & \underline{60.9} & \underline{66.0} & \underline{97.2\%} \\
\midrule
\multicolumn{12}{l}{\textbf{Qwen2.5-Omni-3B}} \\
\midrule
\rowcolor{fullrowcolor} Full Tokens & 100\% & 100\% & 32.9 & 65.3 & 58.4 & 85.0 & 44.1 & 62.6 & 62.2 & 62.6 & 100\% \\
Random & 40\% & 31\% & 28.2 & 60.8 & 54.9 & 73.1 & \underline{42.3} & \underline{61.6} & 57.5 & 60.6 & 94.6\% \\
FastV & 35\% & 37\% & 24.2 & 60.8 & 54.3 & \underline{81.6} & 40.7 & 58.3 & 57.7 & - & 92.8\%$^\dagger$ \\
DyCoke (V\&A) & 35\% & 26\% & \textbf{32.9} & \underline{62.1} & 54.9 & 74.5 & 38.9 & 58.3 & 57.4 & 61.0 & 94.9\% \\
OmniZip & 35\% & 26\% & 28.2 & 58.6 & \underline{57.8} & 80.9 & 41.5 & \textbf{62.1} & 58.7 & \underline{61.9} & 96.6\% \\
\rowcolor{oursrowcolor} DASH (ours) & 35\% & 26\% & \textbf{32.9} & 61.7 & 55.6 & \textbf{83.4} & \textbf{42.4} & 60.0 & \underline{59.9} & \textbf{62.6} & \textbf{98.2\%} \\
\rowcolor{oursrowcolor} DASH (ours) & 25\% & 18\% & 29.0 & \textbf{65.0} & \textbf{59.5} & 81.1 & 39.1 & 58.4 & 58.8 & 61.7 & \underline{96.6\%} \\
\bottomrule
\end{tabular}
}
\end{table*}

\noindent\textbf{Implementation details.}
We implement DASH on Qwen2.5-Omni (7B and 3B) models~\cite{xu2025qwen25omni} using NVIDIA H20 (96GB) GPUs. We use the overall FLOPs ratio as the metric to ensure fair comparison across methods. For video input, we cap the maximum number of frames at 768 for VideoMME and 128 for other datasets. Each time window contains 50 audio tokens and 288 video tokens. For DASH-specific hyperparameters, we set $\tau_a = 0.4$, $C_{\min} = 30$, tri-signal weights $w_b = 0.4$, $w_u = 0.3$, $w_a = 0.3$, and channel selection ratio $0.5$. The reported retention ratios (e.g., 25\%, 35\%) represent target compression levels; in practice, actual per-sample retention may deviate from these targets due to variations in dataset characteristics and individual sample properties (e.g., audio-video token distribution, content complexity). For all experiments, we leverage FlashAttention to reduce memory usage. Evaluation uses deterministic decoding and rule-based multiple-choice parsing.

\subsection{Main Results}
\label{sec:main_results}

We evaluate on Qwen2.5-Omni at two parameter scales (7B and 3B). Following OmniZip, we report baselines at 35\% token retention. DASH is evaluated at both 35\% and 25\% retention to demonstrate its ability to maintain accuracy under more aggressive compression. For VideoMME, we use LMMs-Eval~\cite{zhang2025lmmseval} for evaluation. Following OmniZip, results in \cref{tab:main} are normalized with the baseline model's accuracy set to 100\%.

\noindent\textbf{Comparison with state-of-the-art methods.}
\cref{tab:main} compares DASH with existing methods. The key finding is that \textbf{structure-aware compression enables stable performance at substantially lower token retention}: DASH at 25\% retention achieves 60.9\% AVUT average on the 7B model, competitive with OmniZip at 35\% (60.6\%), despite using only 20\% FLOPs. This 10-percentage-point reduction in token retention with maintained accuracy directly validates our core claim that semantic structure, not positional regularity, should guide compression decisions. On VideoMME, DASH at 25\% matches the full-token baseline (66.0\%), confirming that content-aware chunking preserves information that fixed-size grouping destroys.

Methods ignoring cross-modal structure (Random, FastV) degrade significantly, confirming that temporal window structure is critical for audio-video understanding. The Norm.\ Avg.\ column highlights the overall advantage: DASH at 35\% achieves 98.2\% on both model scales, surpassing OmniZip (97.0\% on 7B, 96.6\% on 3B). The consistency across the 7B and 3B variants suggests that the gains are robust across model scales within the same OmniLLM family.

\begin{table*}[t]
\centering
\caption{\textbf{Comparison of different methods on the WorldSense benchmark.} FLOPs calculation considers only multimodal tokens from audio and video inputs. Best result among token pruning methods is in \textbf{bold}, second best is \underline{underlined}.}
\label{tab:worldsense}
\resizebox{\textwidth}{!}{
\begin{tabular}{l|cc|cccccccc|c}
\toprule
Method & Retained & FLOPs & Tech \& & Culture \& & Daily & Film \& & Perfor- & Games & Sports & Music & Avg. \\
 & Ratio & (T) & Science & Politics & Life & TV & mance & & & & \\
\midrule
\multicolumn{12}{l}{\textbf{Qwen2.5-Omni-7B}} \\
\midrule
\rowcolor{fullrowcolor} Full Tokens & 100\% & 73.2 & 52.4 & 50.1 & 48.5 & 44.6 & 43.8 & 41.6 & 41.6 & 47.3 & 46.8 \\
Random & 55\% & 35.5 & 47.1 & 47.0 & 44.4 & 41.2 & 40.0 & 40.1 & 40.1 & 46.3 & 43.6 \\
FastV & 50\% & 39.3 & \underline{48.8} & 47.4 & 44.2 & \textbf{44.1} & \underline{41.2} & 38.3 & 40.0 & \textbf{46.6} & 44.3 \\
DyCoke (V\&A) & 50\% & 31.9 & 48.4 & \textbf{49.9} & 46.7 & \underline{41.4} & 39.9 & \textbf{40.8} & 40.2 & \underline{46.5} & 44.6 \\
OmniZip & 35\% & 21.4 & \underline{48.8} & \underline{49.5} & \textbf{47.1} & 40.6 & 40.4 & \textbf{40.8} & \underline{40.7} & 45.6 & \underline{44.7} \\
\rowcolor{oursrowcolor} DASH (ours) & 25\% & 14.9 & \textbf{50.0} & 49.2 & \underline{47.0} & 39.3 & \textbf{41.6} & 39.9 & \textbf{41.6} & 45.8 & \textbf{44.9} \\
\midrule
\multicolumn{12}{l}{\textbf{Qwen2.5-Omni-3B}} \\
\midrule
\rowcolor{fullrowcolor} Full Tokens & 100\% & 37.4 & 51.5 & 50.8 & 45.0 & 45.4 & 43.8 & 42.5 & 44.2 & 46.1 & 46.4 \\
Random & 55\% & 17.0 & 48.2 & 46.3 & 40.7 & 41.4 & 38.6 & 40.0 & 41.8 & 43.4 & 42.8 \\
FastV & 50\% & 18.2 & \underline{50.0} & \textbf{50.5} & \underline{44.1} & 43.0 & 40.5 & 41.6 & 41.8 & 42.1 & \underline{44.4} \\
DyCoke (V\&A) & 50\% & 15.1 & 48.1 & 48.5 & 42.3 & \underline{43.3} & 39.7 & \textbf{43.4} & \underline{42.1} & 43.0 & 44.0 \\
OmniZip & 35\% & 9.9 & 48.4 & \underline{49.2} & 41.6 & \textbf{44.3} & \textbf{41.2} & 40.8 & \textbf{43.3} & \textbf{43.6} & 44.1 \\
\rowcolor{oursrowcolor} DASH (ours) & 25\% & 6.7 & \textbf{50.2} & 47.9 & \textbf{44.7} & \underline{43.3} & \textbf{41.2} & \underline{42.5} & 40.9 & \textbf{43.6} & \textbf{44.6} \\
\bottomrule
\end{tabular}
}
\end{table*}

\begin{wrapfigure}[13]{r}{0.5\linewidth}
    \centering
    \vspace{-8mm}
    \includegraphics[width=\linewidth]{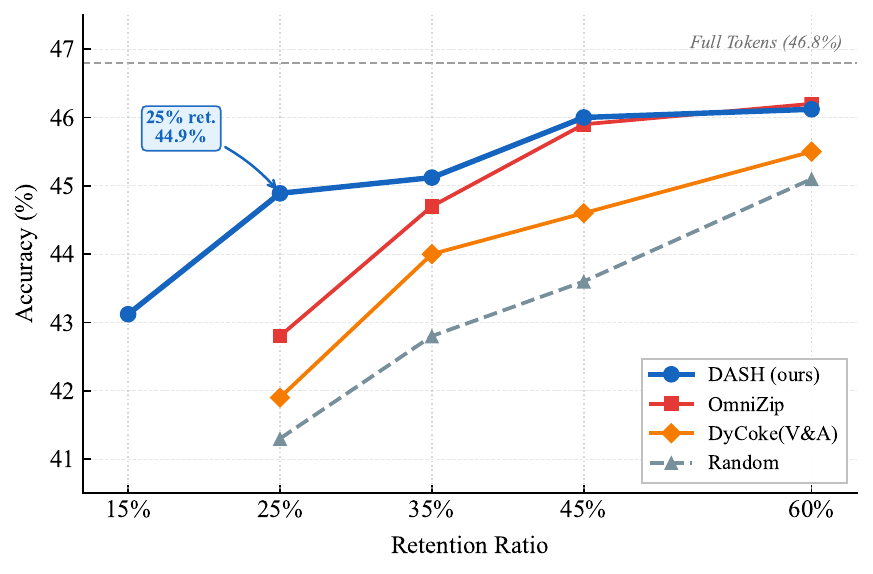}
    \caption{\textbf{Accuracy vs.\ retention ratio on WorldSense (Qwen2.5-Omni-7B).}}
    \label{fig:ratio_sweep}
\end{wrapfigure}

\cref{tab:worldsense} presents per-domain results on WorldSense. DASH at 25\% retention achieves 44.9\% average accuracy on the 7B model, matching or surpassing OmniZip's 44.7\% at 35\% retention while consuming only 14.9T FLOPs versus 21.4T---a 30\% computational saving with improved accuracy. On the 3B model, DASH (44.6\%) likewise outperforms OmniZip (44.1\%) with 32\% fewer FLOPs. Across both scales, DASH leads in domains such as Tech \& Science and Sports, while remaining competitive in others, suggesting that dynamic semantic chunking generalizes well to diverse audio-video content types.

In \cref{fig:ratio_sweep}, we further illustrate the accuracy-retention trade-off of different methods across retention ratios from 15\% to 60\%. DASH maintains a clear advantage at low-to-mid retention, with its 25\% point (44.9\%) already exceeding OmniZip at 35\% (44.7\%) and DyCoke at 45\% (44.6\%). On WorldSense, DASH at 25\% retention reaches the accuracy level that prior methods achieve only around 35--45\% retention. As retention increases, all methods converge toward the Full Tokens ceiling, indicating that the benefit of content-aware compression is most pronounced under aggressive pruning---precisely the regime where efficiency gains matter most for practical deployment.

\subsection{Efficiency Analysis}
\label{sec:efficiency}

\begin{table}[t]
\centering
\caption{\textbf{Inference efficiency on WorldSense.}}
\label{tab:efficiency}
\scriptsize
\renewcommand{\arraystretch}{1.0}
\setlength{\tabcolsep}{2.5pt}
\begin{minipage}[t]{0.49\linewidth}
\centering
\textbf{(a) Qwen2.5-Omni-7B}\\[3pt]
\begin{tabular}{@{}lcccc@{}}
\toprule
Method & Mem.\,$\downarrow$ & Prefill\,$\downarrow$ & Acc.\,$\uparrow$ & Latency\,$\downarrow$ \\
\midrule
\rowcolor{fullrowcolor} Full & 35G & 1.0$\times$ & 46.8 & 1.0$\times$ \\
FastV & OOM & -- & -- & -- \\
DyCoke & 31G & 1.6$\times$ & 44.6 & 1.2$\times$ \\
OmniZip & 25G & 3.4$\times$ & 44.7 & 1.4$\times$ \\
\rowcolor{oursrowcolor} Ours & 26G & 3.5$\times$ & 44.9 & 1.7$\times$ \\
\bottomrule
\end{tabular}
\end{minipage}
\hfill
\begin{minipage}[t]{0.49\linewidth}
\centering
\textbf{(b) Qwen2.5-Omni-3B}\\[3pt]
\begin{tabular}{@{}lcccc@{}}
\toprule
Method & Mem.\,$\downarrow$ & Prefill\,$\downarrow$ & Acc.\,$\uparrow$ & Latency\,$\downarrow$ \\
\midrule
\rowcolor{fullrowcolor} Full & 25G & 1.0$\times$ & 46.4 & 1.0$\times$ \\
FastV & 45G & 1.2$\times$ & 44.4 & 1.1$\times$ \\
DyCoke & 20G & 1.5$\times$ & 44.0 & 1.2$\times$ \\
OmniZip & 16G & 3.3$\times$ & 44.1 & 1.3$\times$ \\
\rowcolor{oursrowcolor} Ours & 16G & 3.8$\times$ & 44.6 & 1.4$\times$ \\
\bottomrule
\end{tabular}
\end{minipage}
\end{table}

\cref{tab:efficiency} reports inference efficiency on WorldSense. At 25\% retention, DASH achieves 3.5$\times$ prefill speedup on the 7B model and 3.8$\times$ on the 3B model, surpassing OmniZip at 35\% (3.4$\times$ and 3.3$\times$) on both scales. End-to-end latency is also reduced to 1.7$\times$ (7B) and 1.4$\times$ (3B), while maintaining competitive accuracy. Critically, the overhead of boundary detection and tri-signal scoring is negligible ($<$40ms), as these operations involve only cosine similarity and element-wise computations on the already-extracted token embeddings. This confirms that structure-aware compression adds minimal computational cost while delivering substantial efficiency gains---a favorable trade-off for practical deployment.

\subsection{Ablation Studies}
\label{sec:ablation}

All ablation experiments are conducted on Qwen2.5-Omni-3B at 25\% retention ratio on the WorldSense benchmark.

\begin{figure}[t]
\centering
\begin{minipage}[t]{0.45\linewidth}
\vspace{-1.3in}
\centering
\footnotesize
\renewcommand{\arraystretch}{1.0}
\setlength{\tabcolsep}{4pt}
\begin{tabular}{@{}lc@{}}
\toprule
Configuration & Acc.\ (\%) \\
\midrule
\rowcolor{fullrowcolor} OmniZip (35\% ret.) & 44.1 \\[2pt]
Static + TSF & 44.4 \\
DSC+ADVS (attn-only) & 44.4 \\
Full DASH & \textbf{44.6} \\
\bottomrule
\end{tabular}
\captionof{table}{\textbf{Ablation on Qwen2.5-Omni-3B (WorldSense, 25\% ret.).} OmniZip at 35\% retention is the reference baseline.}
\label{tab:ablation_unified}
\end{minipage}
\hfill
\begin{minipage}[t]{0.5\linewidth}
\centering
\includegraphics[width=\linewidth]{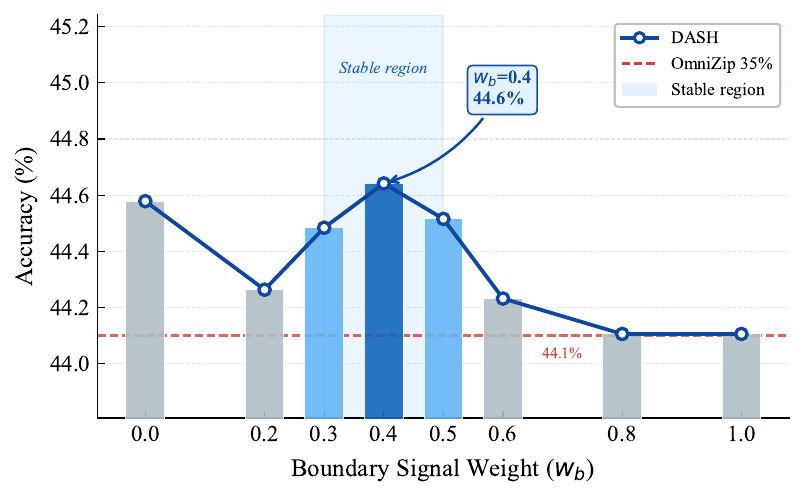}
\captionof{figure}{\textbf{Sensitivity of $w_b$ on WorldSense (3B, 25\% retention).} 
}
\label{fig:signal_weight}
\end{minipage}
\end{figure}

\noindent\textbf{Component ablation and sensitivity analysis.}
\cref{tab:ablation_unified} progressively adds each DASH component. Starting from the baseline with static grouping and attention-only selection, we first add tri-signal fusion alone (Static+TSF), then dynamic chunking with attention-only selection (DSC+ADVS), and finally the full DASH combining all three components. \cref{fig:signal_weight} shows the sensitivity of $w_b$.

From \cref{tab:ablation_unified}, both Static+TSF and DSC+ADVS independently improve over OmniZip at 35\% (44.1\%$\to$44.4\%), confirming that tri-signal fusion and dynamic chunking provide complementary gains of comparable magnitude; combining them in Full DASH further lifts accuracy to 44.6\%. This additive improvement pattern validates our design principle: structure-aware segmentation and multi-signal importance estimation address orthogonal limitations of prior work.

\noindent\textbf{Boundary detection algorithm comparison.}
We evaluate four similarity metrics for audio boundary detection (\cref{tab:boundary_algo_comparison}).

\noindent Cosine similarity achieves the highest accuracy (44.6\%), while dot product (43.6\%), change rate (44.0\%), and random boundaries (43.8\%) perform worse. 
\begin{wraptable}[10]{r}{0.4\linewidth}
    \centering
    \scriptsize
    \renewcommand{\arraystretch}{1.1}
    \setlength{\tabcolsep}{6pt}
    \vspace{-6mm}
    \begin{tabular}{@{}lc@{}}
    \toprule
    Similarity Method & Acc.\ (\%) \\
    \midrule
    Random & 43.8 \\
    Dot Product & 43.6 \\
    Change Rate & 44.0 \\
    Cosine & \textbf{44.6} \\
    \bottomrule
    \end{tabular}
    \caption{\textbf{Boundary detection algorithm comparison on WorldSense (3B, 25\% ret.).}}
    \label{tab:boundary_algo_comparison}
\end{wraptable}
The gap between cosine and dot product confirms the importance of scale invariance in handling feature magnitude variations across audio segments. The consistent advantage over random boundaries indicates that the boundary detection module captures meaningful semantic transitions in audio-visual sequences.

From \cref{fig:signal_weight}, performance is stable for $w_b \in [0.3, 0.5]$, peaking at $w_b{=}0.4$, and degrades for $w_b > 0.5$. The degradation beyond 0.5 suggests that over-emphasizing boundary probability at the expense of content distinctiveness and attention leads to suboptimal selection, validating our default weights.

\section{Conclusion}
\label{sec:conclusion}


We presented DASH, a training-free framework for structure-aware token compression in omnimodal large language models. Rather than treating multimodal tokens as uniformly structured sequences, DASH aligns compression with the intrinsic semantic organization of audio-visual signals. By using audio embeddings as a semantic anchor, DASH detects boundary candidates that approximate the piecewise structure of multimodal sequences and uses them as soft temporal priors for video compression. This design preserves transition-critical information while reducing redundant regions, enabling aggressive compression without disrupting cross-modal coherence. Experiments on AVUT, VideoMME, and WorldSense show that DASH maintains competitive accuracy at substantially lower token retention while significantly improving inference efficiency. These results highlight the importance of aligning compression with semantic structure for efficient omnimodal reasoning.

\section*{Acknowledgments}
This work was supported by the National Natural Science Foundation of China under Grant 62506235.

\bibliographystyle{splncs04}
\bibliography{main}

\appendix

\section{Qualitative Analysis}
\label{sec:qualitative}

\begin{figure}[h]
\centering
\includegraphics[width=\linewidth]{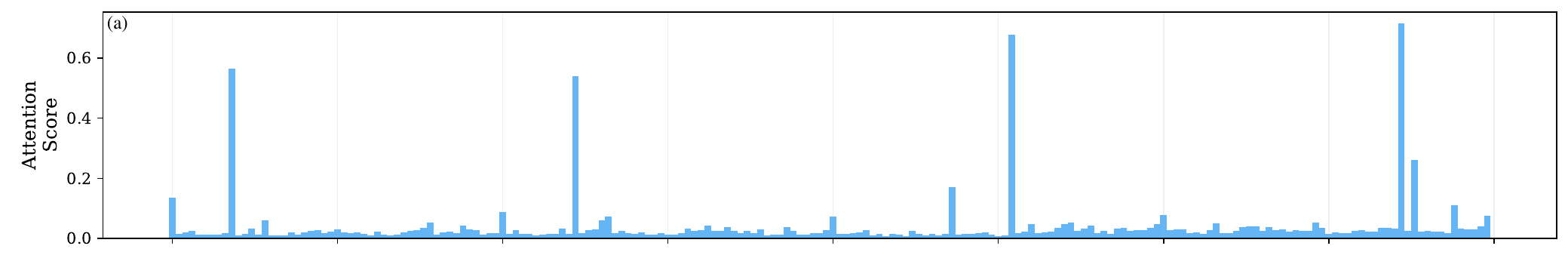}\\[1pt]
\includegraphics[width=\linewidth]{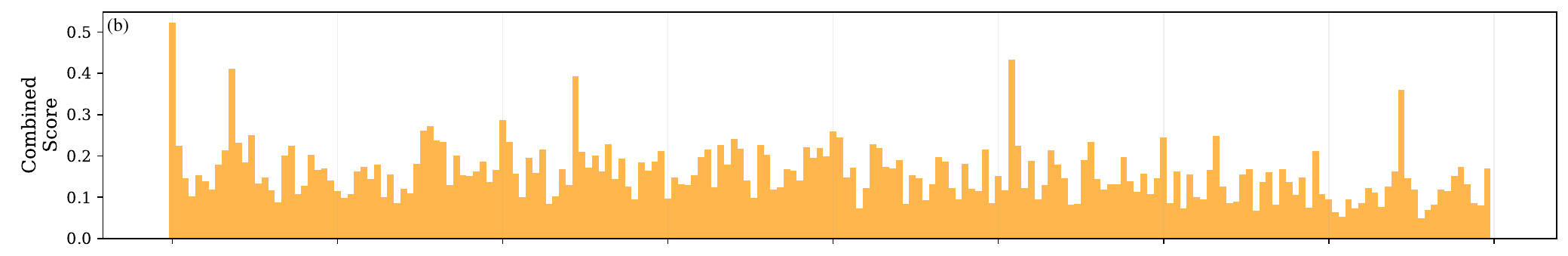}\\[1pt]
\includegraphics[width=\linewidth]{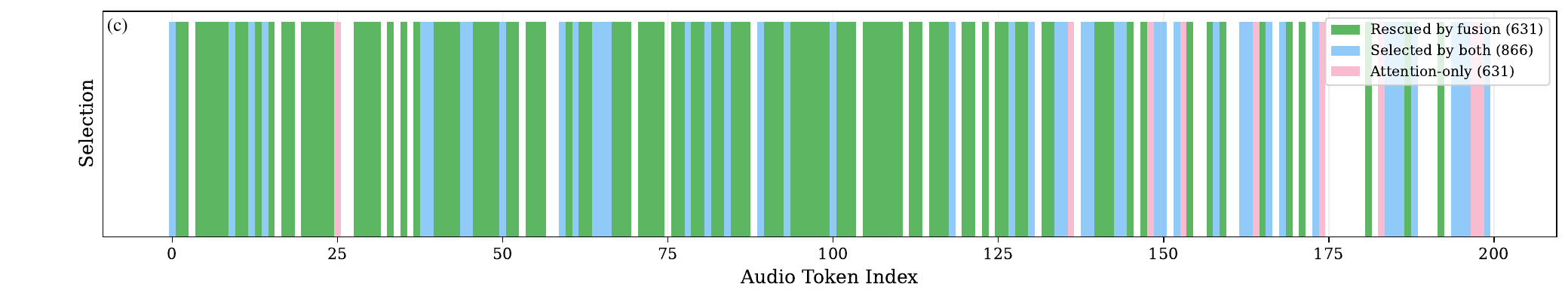}
\caption{\textbf{Token scoring and selection comparison} (reproduced from Fig.~2 in the main paper for reader convenience). (a)~Attention score alone is extremely sparse, concentrating on a few tokens. (b)~Tri-signal combined score produces a balanced importance landscape. (c)~Selection comparison: green = rescued by fusion (631), blue = selected by both (866), pink = attention-only tokens replaced (631). Fusion rescues $\sim$42\% of retained tokens that attention alone would discard.}
\label{fig:token_heatmap2}
\end{figure}

\begin{figure}[t]
\centering
\includegraphics[width=\linewidth]{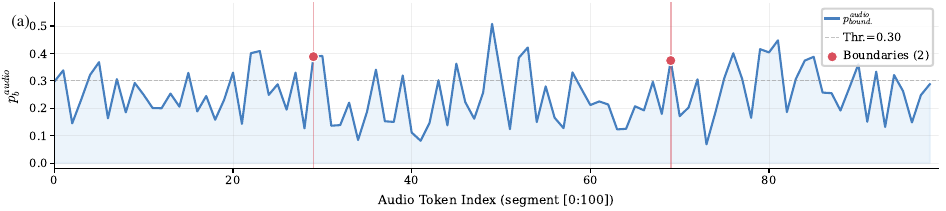}\\[2pt]
\includegraphics[width=\linewidth]{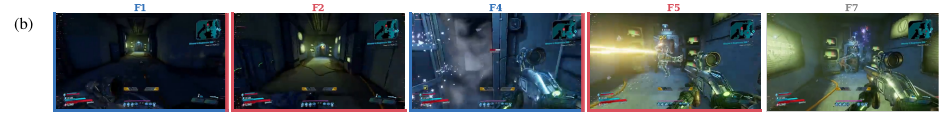}\\[2pt]
\includegraphics[width=\linewidth]{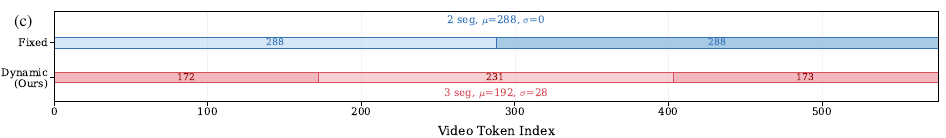}
\caption{\textbf{Boundary detection visualization.} (a)~Boundary probability $p_t^{\text{boundary}}$ over audio tokens (blue curve), with threshold $\tau_a$ (gray dashed) and detected boundaries (red dots). Peaks exceeding the threshold correspond to semantic transitions in the speech stream. (b)~Video frames near projected boundary locations: blue-bordered frames precede and red-bordered frames follow an audio-indicated transition, illustrating that audio boundaries can provide useful temporal priors for visual context. (c)~Fixed grouping (2 equal segments, $\sigma{=}0$) vs.\ dynamic chunking (ours, 3 variable-length segments, $\sigma{=}28$), where $\sigma$ denotes the standard deviation of segment lengths in tokens. Our method adapts segment granularity to content structure rather than imposing uniform partitions.}
\label{fig:boundary_vis}
\end{figure}

\noindent\textbf{Boundary detection visualization.}
\cref{fig:boundary_vis} visualizes audio-driven dynamic segmentation on a representative segment. (a)~plots the raw boundary probability $p_t^{\text{boundary}}$ over audio tokens; detected boundaries (red dots) correspond to positions where the probability exceeds the threshold (gray dashed line), indicating speech pauses or topic shifts. (b)~displays the video frames surrounding each projected boundary, where blue-bordered frames precede and red-bordered frames follow an audio-indicated transition, illustrating the temporal prior used by DASH rather than assuming exact audio-visual boundary equality. (c)~contrasts fixed-size grouping (2 equal segments, $\sigma{=}0$) with our dynamic chunking (3 variable-length segments, $\sigma{=}28$): our method introduces an additional segment boundary at a semantically meaningful position, adapting granularity to content structure rather than imposing uniform partitions.

\noindent\textbf{Token importance heatmap.}
\cref{fig:token_heatmap} visualizes the attention sparsity discussed in the Introduction Section on the first 200 audio tokens. (a)~shows that attention scores are dominated by near-zero values with isolated peaks---a heavy-tailed distribution that makes attention-only selection brittle under aggressive compression. (b)~shows tri-signal fusion produces a smoother importance landscape by filling gaps attention alone misses, rescuing structurally critical tokens at semantic boundaries. (c)~quantifies the impact: among top-50\% retained tokens, 631 are rescued by fusion, 866 are shared, and 631 attention-only tokens are replaced---a 42\% turnover confirming that fusion fundamentally reshapes selection. This high turnover rate explains why DASH maintains accuracy at 25\% retention where attention-only methods degrade: the rescued tokens are precisely those that preserve narrative continuity and cross-modal coherence.

These qualitative results corroborate the quantitative findings in our experiments: fusion rescues structurally critical tokens that sparse attention would discard, with the gap widening under aggressive compression. Together, \cref{fig:boundary_vis} and \cref{fig:token_heatmap} provide visual evidence for the two core innovations of DASH---dynamic semantic chunking adapts to content structure, and tri-signal fusion captures importance dimensions that attention alone misses.

\section{Algorithm Overview}
\label{sec:algorithm}

Algorithm~\ref{alg:dash} presents the complete DASH pipeline. Given audio tokens $\mathbf{A}\in\mathbb{R}^{N_a\times D}$, video tokens $\mathbf{V}\in\mathbb{R}^{N_v\times D}$ (with $N_v = F \times K$, where $F$ is the number of frames and $K$ is the number of tokens per frame), and audio attention scores, DASH proceeds in four stages: (1)~audio boundary detection via cosine similarity (Sec.~3.2 in the main paper), (2)~audio-to-video boundary projection as a soft temporal prior with strength-based refinement (Sec.~3.3), (3)~tri-signal fusion for audio token selection (Sec.~3.4), and (4)~adaptive video compression with boundary-neighborhood retention (Sec.~3.5). All operations are training-free and require only element-wise or pairwise computations on the already-extracted encoder embeddings.

\begin{algorithm}[!h]
\caption{DASH: Dynamic Audio-Driven Semantic Chunking}
\label{alg:dash}
\small
\begin{algorithmic}[1]
\REQUIRE Audio tokens $\mathbf{A}\!=\!\{a_t\}_{t=1}^{N_a}$, video tokens $\mathbf{V}\!=\!\{v_t\}_{t=1}^{N_v}$, attention scores $\texttt{attn}$, audio compression ratio $\rho_a$, video compression ratio $\rho_v$, threshold $\tau_a$, minimum chunk size $C_{\min}$, tokens per frame $K$
\ENSURE Retained audio mask $\mathbf{m}_a$, retained video mask $\mathbf{m}_v$

\STATE \textbf{// Stage 1: Dynamic Semantic Chunking (Sec.~3.2 in main paper)}
\STATE Compute cosine similarity: $\text{sim}_t \leftarrow \frac{\langle a_{t-1},\, a_t \rangle}{\|a_{t-1}\|\,\|a_t\|}$ for $t = 2, \ldots, N_a$
\STATE Compute boundary probability: $p_t^{\text{bnd}} \leftarrow \text{clip}((1 - \text{sim}_t)/2,\; 0,\; 1)$
\STATE $\mathcal{B}_a \leftarrow \{0\}$; \quad $t_{\text{last}} \leftarrow 0$
\FOR{$t = 2$ \TO $N_a$}
    \IF{$\text{sim}_t < \tau_a$ \AND $t - t_{\text{last}} \geq C_{\min}$}
        \STATE $\mathcal{B}_a \leftarrow \mathcal{B}_a \cup \{t\}$; \quad $t_{\text{last}} \leftarrow t$
    \ENDIF
\ENDFOR
\STATE $\mathcal{B}_a \leftarrow \mathcal{B}_a \cup \{N_a\}$ \COMMENT{Append end position}

\STATE \textbf{// Stage 2: Audio-Guided Video Segmentation (Sec.~3.3)}
\STATE Project: $b_i^v \leftarrow \lfloor b_i^a \cdot N_v / N_a \rfloor$ for each $b_i^a \in \mathcal{B}_a$
\STATE Sort inner boundaries $\{b_i^v\}_{i=1}^{|\mathcal{B}_v|-2}$ by strength $p_{b_i^a}^{\text{bnd}}$ in descending order
\STATE $\mathcal{B}_v^* \leftarrow \{0,\; N_v\}$
\FOR{each inner boundary $b_i^v$ in strength order}
    \IF{$(b_i^v - b_{\text{left}}) \geq 2K$ \AND $(b_{\text{right}} - b_i^v) \geq 2K$}
        \STATE Insert $b_i^v$ into $\mathcal{B}_v^*$
    \ENDIF
\ENDFOR

\STATE \textbf{// Stage 3: Tri-Signal Fusion Token Scoring (Sec.~3.4)}
\STATE $s_t^{\text{bnd}} \leftarrow p_t^{\text{bnd}} \,/\, (\max_j p_j^{\text{bnd}} + \epsilon)$ \COMMENT{Boundary signal}
\STATE Select low-variance channels: $\tilde{\mathbf{A}} \leftarrow \text{ChannelSelect}(\mathbf{A}, 0.5)$
\STATE $\hat{\mathbf{A}} \leftarrow \ell_2\text{-normalize}(\tilde{\mathbf{A}})$; \quad $\mathbf{c} \leftarrow \text{mean}(\hat{\mathbf{A}})$
\STATE $g_t \leftarrow \sum_{\alpha \in \mathcal{A}} \exp(-\|\hat{a}_t - \mathbf{c}\|^2 / 2\alpha)$ \COMMENT{Multi-scale Gaussian}
\STATE $s_t^{\text{uniq}} \leftarrow 1 - g_t / (\max_j g_j + \epsilon)$ \COMMENT{Uniqueness signal}
\STATE $s_t^{\text{attn}} \leftarrow \texttt{attn}_t \,/\, (\max_j \texttt{attn}_j + \epsilon)$ \COMMENT{Attention signal}
\STATE $s_t \leftarrow 0.4 \cdot s_t^{\text{bnd}} + 0.3 \cdot s_t^{\text{uniq}} + 0.3 \cdot s_t^{\text{attn}}$
\STATE $\mathbf{m}_a \leftarrow \text{top-}\lfloor (1-\rho_a) \cdot N_a \rfloor$ tokens by $s_t$

\STATE \textbf{// Stage 4: Boundary-Aware Video Compression (Sec.~3.5)}
\FOR{each video segment $s$ defined by $\mathcal{B}_v^*$}
    \STATE Compute audio retention $\bar{m}_a^{(s)}$ for the corresponding audio segment
    \STATE $\rho_v^{(s)} \leftarrow \text{clip}(\rho_v + 0.1 \cdot (0.5 - \bar{m}_a^{(s)}),\; 0.1,\; 0.95)$
    \FOR{each frame $f$ in segment $s$}
        \STATE $r_f \leftarrow 1 - \rho_v^{(s)}$ \COMMENT{Base frame retention}
        \IF{frame $f$ is at a boundary position}
            \STATE $r_f \leftarrow r_f + (1 - r_f) \cdot 0.3 \cdot p_f^{\text{bnd}}$ \COMMENT{Boundary protection}
        \ENDIF
        \STATE Apply interleaved spatial-temporal pruning with retention $r_f$
    \ENDFOR
\ENDFOR
\RETURN $\mathbf{m}_a$, $\mathbf{m}_v$
\end{algorithmic}
\end{algorithm}

\section{Implementation Details}
\label{sec:impl_details}

\noindent\textbf{Hyperparameter summary.}
Table~\ref{tab:hyperparams} lists all DASH-specific hyperparameters used across experiments. These values are fixed for all benchmarks and both model scales (7B and 3B) unless otherwise noted.

\begin{table}[!htb]
\centering
\caption{\textbf{DASH hyperparameters.} All values are fixed across benchmarks and model scales.}
\label{tab:hyperparams}
\small
\renewcommand{\arraystretch}{1.15}
\begin{tabular}{@{}llll@{}}
\toprule
Symbol & Value & Eq. & Description \\
\midrule
$\tau_a$ & 0.4 & (3) & Cosine similarity threshold for boundary detection \\
$C_{\min}$ & 30 & (3) & Minimum chunk size ($\approx$1\,s of audio) \\
$2K$ & $2 \times K$ & (5) & Minimum video segment size for ISTC \\
$w_b$ & 0.4 & (10) & Boundary probability weight in tri-signal fusion \\
$w_u$ & 0.3 & (10) & Uniqueness weight in tri-signal fusion \\
$w_a$ & 0.3 & (10) & Attention weight in tri-signal fusion \\
$\mathcal{A}$ & $\{2^k\}_{k=-3}^{1}$ & (7) & Multi-scale Gaussian bandwidth set \\
Channel ratio & 0.5 & -- & Fraction of low-variance channels retained \\
$\lambda_r$ & 0.1 & (11) & Adaptation strength for video compression ratio \\
$[\rho_v^{\min}, \rho_v^{\max}]$ & $[0.1, 0.95]$ & (11) & Clamping range for adaptive video compression \\
\bottomrule
\end{tabular}
\end{table}

\noindent\textbf{Time-window structure.}
Following OmniZip~\cite{tao2025omnizip}, the Qwen2.5-Omni model organizes multimodal inputs into time windows. Each window contains 50 audio tokens and 288 video tokens. DASH operates independently on each time window, detecting boundaries and performing compression within the window's audio and video token sequences.

\noindent\textbf{Evaluation protocol.}
We evaluate on three benchmarks:
\begin{itemize}
    \item \textbf{AVUT}~\cite{yang2025avut}: An audio-centric video understanding benchmark with six subtasks---event localization (EL), object matching (OM), OCR matching (OR), information extraction (IE), content counting (CC), and character matching (CM). We report per-task accuracy and the overall average.
    \item \textbf{VideoMME}~\cite{fu2025videomme}: A general video understanding benchmark evaluated using LMMs-Eval~\cite{zhang2025lmmseval}. We report the ``without subtitle'' setting to focus on audio-visual reasoning. The maximum number of input frames is capped at 768.
    \item \textbf{WorldSense}~\cite{hong2025worldsense}: A benchmark for joint audio-video understanding across eight domains (Tech \& Science, Culture \& Politics, Daily Life, Film \& TV, Performance, Games, Sports, Music). The maximum number of input frames is capped at 128.
\end{itemize}

\noindent\textbf{Baseline reproduction.}
For fair comparison, we reproduce results for OmniZip and DASH under identical settings (same model checkpoints, input preprocessing, and evaluation scripts). Results for Random, FastV~\cite{chen2024fastv}, and DyCoke~\cite{tao2025dycoke} are taken directly from OmniZip~\cite{tao2025omnizip} as reported in their paper.

\end{document}